\documentclass[12pt]{article}

\newcommand{\be}{\begin{equation}}
\newcommand{\ee}{\end{equation}}
\newcommand{\bi}[1]{\bibitem{#1}}

\usepackage{epsfig,amsmath,amssymb,graphics,graphicx}

\begin{document}

\begin{center}
{\it Journal of Mathematical Physics 49 (2008) 102112}
\vskip 5 mm

{\Large \bf Weyl Quantization of Fractional Derivatives}
\vskip 5 mm

{\large \bf Vasily E. Tarasov}\\
\vskip 3mm

{\it Skobeltsyn Institute of Nuclear Physics, \\
Moscow State University, Moscow 119991, Russia \\
E-mail: tarasov@theory.sinp.msu.ru }
\end{center}

\vskip 7 mm

\begin{abstract}
The quantum analogs of the derivatives 
with respect to coordinates $q_k$ and momenta $p_k$
are commutators with operators $P_k$ and $Q_k$. 
We consider quantum analogs of  
fractional Riemann-Liouville and Liouville derivatives.
To obtain the quantum analogs of  fractional Riemann-Liouville derivatives,
which are defined on a finite interval of the real axis, 
we use a representation of these derivatives for analytic functions.
To define a quantum analog of the fractional Liouville derivative,
which is defined on the real axis, 
we can use the representation of the Weyl quantization
by the Fourier transformation.
\end{abstract}





\newpage

\section{Introduction}

It is well known that the derivatives with respect to coordinates $q_k$ 
and momenta $p_k$ can be represented as Poisson brackets by the equations
\[ \frac{\partial}{\partial q_k} A(q,p) = -\{p_k, A(q,p)\} , \]
\[ \frac{\partial}{\partial p_k} A(q,p) = \{q_k, A(q,p)\}  \]
for continuously differentiable functions $A(q,p) \in C^1(\mathbb{R}^{2n})$.
Quantum analogs of these Poisson brackets are self-adjoint commutators.
The Weyl quantization $\pi$ gives
\[ \pi \Bigl( \{q_k, A(q,p)\} \Bigr)= \frac{1}{i\hbar} [\pi(q_k), \pi(A) \} ] , \]
\[ \pi \Bigl( \{p_k, A(q,p)\} \Bigr)= \frac{1}{i\hbar} [\pi(p_k), \pi(A) \} ] , \]
where $[A,B]=AB-BA$.
As a result, we have that
\be \label{commutators}
 L^{-}_{P_k}=-(1/i\hbar) [\pi(q_k), \ . \ ] , \quad
-L^{-}_{Q_k}=-(1/i\hbar) [\pi(q_k), \ . \ ]   
\ee
can be consider as quantum analogs of derivatives 
$\partial/\partial q_k$ and $\partial/\partial p_k$.
Then a quantum analog of a derivative of integer order $n$
can be defined by the products of $L^{-}_{P_k}$ and $L^{-}_{Q_k}$.
For example, a quantum analog of $\partial^2/ \partial q_k \partial p_l$
has the form
$-L^{-}_{P_k}L^{-}_{Q_l}=-(1/\hbar)^2 [P_k,[Q_l, \ . \ ]]$.
An analog of $\partial^n/\partial q^n_k$ is $(-1)^n (L^{-}_{P_k})^n$. 

The theory of derivatives of noninteger order 
goes back to Leibniz, Liouville, Grunwald, Letnikov and 
Riemann \cite{KST,OS,SKM,MR,Podlubny},
and the theory has found many applications in recent studies in physics 
(see, for example \cite{Zaslavsky1,GM,H,Zaslavsky2,MS,SATM} 
and references therein).
The fractional derivative has different definitions \cite{SKM,KST}, 
and exploiting any of them depends on the kind of 
problems, initial (boundary) conditions, and 
specifics of the considered physical processes.
The classical definition are the so-called Riemann-Liouville and
Liouville derivative \cite{KST}.
These fractional derivatives are defined by the same equations 
on a finite interval of $\mathbb{R}$ and of the real axis correspondently. 
Note that the Caputo and Riesz derivatives 
can be represented \cite{KST,SKM} thought the Riemann-Liouville and Liouville derivatives.
Therefore quantum analogs of the fractional Riemann-Liouville 
and Liouville derivatives
allow us to derive quantum analogs for Caputo and Riesz derivatives.

The quantum analogs of the derivatives 
$\partial/\partial q_k$ and $\partial/\partial p_k$
are commutators (\ref{commutators}). 
What are quantum analogs of  
fractional Riemann-Liouville and Liouville derivatives?
To obtain the quantum analogs of  fractional Riemann-Liouville derivatives,
which are defined on a finite interval of $\mathbb{R}$, 
we can use a representation of these derivatives for analytic functions. 
In this representation the Riemann-Liouville derivative
is a series of derivatives of integer order.
It allows us to use the correspondence between the integer derivatives
and the self-adjoint commutators.
To define a quantum analog of the fractional Liouville derivative,
which is defined on the real axis $\mathbb{R}$, 
we can use the representation of the Weyl quantization
by the Fourier transformation.

\section{Quantization of fractional Riemann-Liouville derivative}

The fractional derivative $_aD^{\alpha}_{x}$ on $[a,b]$
in the Riemann-Liouville form is defined by the equation
\[ _0D^{\alpha}_{x}A(x)=\frac{1}{\Gamma (m-\alpha)}
\frac{d^m}{d x^m}  
\int^x_{a} \frac{A(y) dy}{(x-y)^{\alpha-m+1}} , \]
where $m$ is the first whole number greater than or equal to $\alpha$.
To simplify our equation, we use $a=0$.
The derivative of powers $n$ of $x$ is
\be \label{xn} _0D^{\alpha}_x x^n=
\frac{\Gamma(n+1)}{\Gamma(n+1-\alpha)} x^{n-\alpha} ,\ee
where $n \ge 1$ and $\alpha \ge 0$. 
Here $\Gamma(z)$ is a gamma function.

Let $A(x)$ be an analytic function for $x \in (0,b)$.
The fractional Riemann-Liouville derivative 
on the interval $[0,b]$ can be presented 
(see Lemma 15.3 in Ref. \cite{SKM}) in the form
\[ _0D^{\alpha}_x A (x)=\sum^{\infty}_{n=0} a(n,\alpha) 
x^{n-\alpha} \, \frac{d^n A(x)}{dx^n} , \]
where
\[ a(n,\alpha) =\frac{\Gamma(\alpha+1)}{\Gamma(n+1)
\Gamma(\alpha-n+1) \Gamma(n-\alpha+1)} . \]

If $A(q,p)$ is an analytic function on $M \subset \mathbb{R}^{2n}$,
then we can define the fractional derivatives
\be \label{Fder1} 
_0D^{\alpha}_{q_k} A(q,p)=\sum^{\infty}_{n=0} a(n,\alpha) \,
q^{n-\alpha}_k \, \frac{\partial^n }{ \partial q^n_k} A(q,p), \ee
\be \label{Fder2}
_0D^{\alpha}_{p_k} A(q,p)=\sum^{\infty}_{n=0} a(n,\alpha) \,
p^{n-\alpha}_k \, \frac{\partial^n }{ \partial p^n_k} A(q,p), \ee
where $k=1,...,N$. 
Using the operators 
\[ L^{+}_{q_k}A(q,p)=q_k A(q,p) , \quad
L^{+}_{p_k}A(q,p)=p_k A(q,p) , \]
\[ L^{-}_{q_k}A(q,p)= \frac{\partial A(q,p)}{\partial p_k} , \quad
L^{-}_{p_k}A(q,p)=-\frac{\partial A(q,p)}{\partial q_k} , \]
equations (\ref{Fder1}) and (\ref{Fder2}) can be rewritten in the forms 
\[ _0D^{\alpha}_{q_k} A(q,p)=\sum^{\infty}_{n=0} a(n,\alpha) \,
(L^{+}_{q_k})^{n-\alpha} \, (-L^{-}_{p_k})^n A(q,p), \]
\[ _0D^{\alpha}_{p_k} A(q,p)=\sum^{\infty}_{n=0} a(n,\alpha) \,
(L^{+}_{p_k})^{n-\alpha} \, (L^{-}_{q_k})^n A(q,p). \]
As a result, the fractional derivatives are defined by
\be \label{Fder3} 
_0D^{\alpha}_{q_k} =\sum^{\infty}_{n=0} a(n,\alpha) \,
(L^{+}_{q_k})^{n-\alpha} \, (-L^{-}_{p_k})^n , \ee
\be \label{Fder4}
_0D^{\alpha}_{p_k} =\sum^{\infty}_{n=0} a(n,\alpha) \,
(L^{+}_{p_k})^{n-\alpha} \, (L^{-}_{q_k})^n . \ee

The Weyl quantization $\pi$ of $q_k$ and $p_k$ gives the operators
\[ Q_k=\pi(q_k), \quad P_k=\pi(p_k) . \]
The Weyl quantization of the operators 
$L^{\pm}_{q^k}$ and $L^{\pm}_{p^k}$ is defined \cite{PLA2001,kn3} by the equation
\[ \pi_W(L^{+}_{q_k})=L^{+}_{Q_k}  , \quad 
\pi_W(L^{-}_{q_k})=L^{-}_{Q_k} , \]
\[ \pi_W(L^{+}_{p_k})=L^{+}_{P_k}  , \quad 
\pi_W(L^{-}_{p_k})=L^{-}_{P_k} , \]
where
\be \label{34LL1} 
L^{+}_{Q} A=\frac{1}{2}(Q A+ A Q) ,\quad
L^{+}_{P} A=\frac{1}{2}(P A+ A P) , \ee
\be \label{34LL2}
L^{-}_{Q} A=\frac{1}{i \hbar}(QA - AQ) , \quad
L^{-}_{P} A=\frac{1}{i \hbar}(PA - AP) . \ee

As a result, the quantization of the fractional derivatives 
(\ref{Fder3}) and (\ref{Fder4}) gives the superoperators
\be \label{Fder5} 
_0{\cal D}^{\alpha}_{Q_k} = \pi(\, _0D^{\alpha}_{q_k})=
\sum^{\infty}_{n=0} a(n,\alpha) \,
(L^{+}_{Q_k})^{n-\alpha} \, (-L^{-}_{P_k})^n , \ee
\be \label{Fder6}
_0{\cal D}^{\alpha}_{P_k} = \pi(\, _0D^{\alpha}_{p_k})=
\sum^{\infty}_{n=0} a(n,\alpha) \,
(L^{+}_{P_k})^{n-\alpha} \, (L^{-}_{Q_k})^n . \ee
Note that a superoperator is a rule that assigns to each operator 
exactly one operator (see, for example Ref. \cite{kn3}).
Equations (\ref{Fder5}) and (\ref{Fder6}) can be considered as
definitions of the fractional derivation superoperators 
on an operator space (for example, 
on rigged operator Hilbert space \cite{kn3}). 

It is not hard to prove that
\[ _0{\cal D}^{\alpha}_{Q} Q^n=
\frac{\Gamma(n+1)}{\Gamma(n+1-\alpha)} Q^{n-\alpha} ,\quad
 _0{\cal D}^{\alpha}_{P} P^n=
\frac{\Gamma(n+1)}{\Gamma(n+1-\alpha)} P^{n-\alpha} ,\]
where $n \ge 1$, and $\alpha \ge 0$.

\section{Quantization of fractional Liouville derivative}

$ \quad \ $ 
We are reminded of the formula for the Fourier transform $ \tilde A(a)$
of some function $A(x)$:
\[ \tilde A(a)={\cal F}\{A(x)\}=
\frac{1}{(2\pi)^{1/2}} \int_{\mathbb{R}} dx \, A(x) \, \exp \{-iax\} , \]
which is valid for all $A(x)$ with
\[ \int_{\mathbb{R}} dx \, |A(x)| < \infty . \]
If we require
\[ \|A(x)\|_2=\int_{\mathbb{R}} dx \, |A(x)|^2 < \infty , \]
then the Parseval formula $\|\tilde A\|_2=\|A\|_2$ holds. 

Let ${\cal F}$ be an extension of this Fourier transformation
to a unitary isomorphism on $L_2(\mathbb{R})$. 
We define the operators
\[ {\cal L}={\cal F}^{-1} L(a) {\cal F} . \]
It is well defined if the function $L(a)$ is measurable. 
These operators form a commutative algebra.
Let ${\cal L}_1$ and ${\cal L}_2$ be operators associated
with the functions $L_1(a)$ and $L_2(a)$.
If ${\cal L}_{12}$ is an operator associated with 
$L_{12}(a)=L_1(a)L_2(a)$, then  
\[ {\cal L}_{12}={\cal L}_1{\cal L}_2={\cal L}_2{\cal L}_1 . \]

As a result, we may present explicit formulas for 
a fractional derivative.
If the Fourier transforms exists,
then the operator $D^{\alpha}_x$ is
\[ D^{\alpha}_x A(x)={\cal F}^{-1} (i a)^{\alpha} \tilde A(a) =
{\cal F}^{-1} (i a)^{\alpha} {\cal F} A(x), \]
where
\[ (ia)^{\alpha}=|a|^{\alpha} \exp\Bigl( \frac{\pi \alpha}{2} sgn(a)\Bigr) . \]
In this paper, we use the following assumption 
that is usually used \cite{Yosida,SKM}. 
The branch of $z^{\alpha}$ is so taken that $Re(z^{\alpha})>0$ 
for $Re(z)>0$. 
This branch is a one-valued function in the $z$-plane 
cut along the negative real axis.

For $A(x)\in L_2(\mathbb{R})$, we have the integral representation
\be \label{Kem1}
D^{\alpha}_x A(x)=
\frac{1}{2\pi} \int_{\mathbb{R}^2} da dx' \, (ia)^{\alpha} \, 
A(x') \, \exp \{ia(x-x')\} . \ee
Some elementary manipulations lead to the well-known Riemann-Liouville
integral representation
\[ D^{\alpha}_x A(x)=\, _{-\infty}D^{\alpha}_x A(x)=
\frac{1}{\Gamma(m-\alpha)}
\frac{d^m}{d x^m} \int^{x}_{-\infty} dx' \,  
\frac{A(x')}{(x-x')^{\alpha+1-m}} . \]
This form cannot be used to quantization.
For the Weyl quantization, we consider the representation (\ref{Kem1}). 

Let $A(q,p)$ be a function of $L_2(\mathbb{R}^2)$ on the phase space
$\mathbb{R}^2$.
Then equation (\ref{Kem1}) can be presented in the form
\be \label{Kem2}
D^{\alpha}_q D^{\beta}_p A(q,p)=
\int_{\mathbb{R}^4} \frac{da db \, dq' dp'}{(2\pi \hbar)^2} 
\, (ia)^{\alpha} \, (ib)^{\beta} \,
A(q',p') \, e^{\frac{i}{\hbar} \left( a(q-q')+b(p-p')\right) }. \ee
Using the Weyl quantization \cite{Berez,kn3} of $A(q,p)$ in the form
\[ A(Q,P)=\pi(A(q,p)) = \int_{\mathbb{R}^4} 
\frac{da db \, dq dp}{(2\pi \hbar)^2} \, 
A(q,p) \, e^{\frac{i}{\hbar} \left( a(Q-qI)+b(P-pI)\right)} , \]
we obtain the following result of the Weyl quantization of (\ref{Kem2}):
\[ D^{\alpha}_Q D^{\beta}_P A(Q,P)=
\pi\Bigr(D^{\alpha}_q D^{\beta}_p A(q,p)\Bigl)= \]
\be \label{Kem3}
=\frac{1}{(2\pi \hbar)^2} \int_{\mathbb{R}^4} da db \, dq dp \, 
(ia)^{\alpha} \, (ib)^{\beta} \,
A(q,p) \, \exp \frac{i}{\hbar} \Bigl( a(Q-qI)+b(P-pI)\Bigr) . \ee
This equation can be considered as a definition of 
$D^{\alpha}_Q$ and $D^{\beta}_P$ on a set of quantum observables.

Note that the general quantization \cite{BalJen,kn3} of $A(q,p)$ is defined by
\[ A_F(Q,P)= \int_{\mathbb{R}^4} 
\frac{da db \, dq dp }{(2\pi \hbar)^2}\, F(a,b) \,
A(q,p) \, \exp \frac{i}{\hbar} \Bigl( a(Q-qI)+b(P-pI)\Bigr) . \]
For the Weyl quantization, $F(a,b)=1$. 
If $F(a,b)=\cos(ab/2\hbar)$, then we have the Rivier quantization \cite{BalJen}. 
Equation (\ref{Kem3}) can be considered as a general quantization
of $A(q,p)$ with the function
\[ F(a,b)= (ia)^{\alpha} \, (ib)^{\beta} . \]
This function has zeros on the real $a$, $b$ axis.
It is clear that there is no dual operator basis.

\section{Conclusion}

The quantum dynamics can be described by superoperators.
A superoperator is a map that assigns to each operator exactly one operator. 
The natural description of the motion is in terms of the 
infinitesimal change in the system.
In the equations of quantum systems the infinitesimal generators 
are defined by some derivation superoperators. 
A derivation is a linear map ${\cal D}$, 
which satisfies the Leibnitz rule
${\cal D}(AB)=({\cal D}A)B+ A({\cal D}B)$ for all operators $A$ and $B$.
It is known that the superoperators ${\cal D}_P=(1/i\hbar)[Q, \ . \ ]$ 
and ${\cal D}_Q=(-1/i\hbar)[P, \ . \ ]$, 
which are used in equations of motion, are derivations of observables.
For example, the quantum harmonic oscillator with the Hamiltonian
$H=(1/2m)P^2+(m \omega^2/2)Q^2$ is described by 
the equation of motion $d A/dt={\cal L}A$ with the infinitesimal generator
${\cal L}=-(1/m)L^{+}_P {\cal D}_Q+ m \omega^2 L^{+}_Q {\cal D}_P$. 
We can consider fractional derivatives on
a set of quantum observables 
as fractional powers ${\cal D}^{\alpha}_Q$ and ${\cal D}^{\alpha}_P$ 
of derivative superoperators ${\cal D}_Q$ and ${\cal D}_P$. 
Note that a fractional generalization of 
the Heisenberg equation is suggested in Ref. \cite{PLA2008}.

In this paper, a fractional generalization of the derivative superoperators 
on a set of quantum observables is suggested.
A fractional power $\alpha$ of the superoperator ${\cal D}$ 
can be considered as a parameter to describe a measure 
of "screening" of environment.
There exist the following special cases of the measure: 
(1) the absence of the environmental influence ($\alpha=0$); 
(2) the complete environmental influence ($\alpha=1$); and
(3) the powerlike environmental influence ($0<\alpha<1$). 
As a result, a physical interpretation of fractional powers of 
the derivative superoperator ${\cal D}_Q$ and ${\cal D}_P$ 
can be a one-parameter description of 
a screening of interaction with environment.

Using the Weyl quantization and
the representation of fractional derivative 
for analytic functions (see Lemma 15.3 in Ref. \cite{SKM}) 
quantum analogs of the Riemann-Liouville and Liouville derivatives
can be obtained.
The Caputo and Riesz derivatives can be represented \cite{KST,SKM} 
thought the Riemann-Liouville and Liouville derivatives.
Therefore quantum analogs of the fractional Riemann-Liouville 
and Liouville derivatives allow us to derive 
quantum analogs for the Caputo and Riesz derivatives.
Note that a quantum analog of a fractional derivative can be 
considered as a fractional power of a self-adjoint commutator \cite{PLA2008}. 
Quantum analogs of fractional derivatives 
can give us a notion that allows one to consider quantum processes 
that are described by fractional differential equations
at classical level (see, for example Ref. \cite{JMP2006} and references therein). 

Note that the fractional equations of motion describe 
an anomalous diffusion \cite{Zaslavsky1,Zaslavsky2,MS,MK1}.
It is known that the quantum Markovian equations (the Lindblad equations) 
are used to describe Brownian motion of quantum systems \cite{Lind2}. 
We can assume that fractional power of derivative superoperators 
in the Lindblad equation can be used to describe 
anomalous processes and continuous time random walks in quantum systems.


\end{document}